\documentclass[prb, twocolumn, floatfix]{revtex4}
\usepackage{graphicx}
\usepackage{epstopdf}

\begin{document}

\title{Quantum feedback control of a superconducting qubit: Persistent Rabi oscillations}
\author{R. Vijay$^1$, C. Macklin$^1$, D. H. Slichter$^1$, S. J. Weber$^1$, K. W. Murch$^1$, R. Naik$^1$, \\
 A. N. Korotkov$^2$,  I. Siddiqi$^{1}$}
\affiliation{$^1$Quantum Nanoelectronics Laboratory, Department of Physics, University of California, Berkeley CA 94720}
\affiliation{$^2$Department of Electrical Engineering, University of California, Riverside, CA 92521}

\date{\today}

\maketitle

\date{\today}

The act of measurement bridges the quantum and classical worlds by projecting a superposition of possible states into a single, albeit probabilistic, outcome. The time-scale of this ``instantaneous" process can be stretched using weak measurements \cite{qmcontrol_book,qnoise_book} so that it takes the form of a gradual random walk towards a final state. Remarkably, the interim measurement record is sufficient to continuously track and steer the quantum state using feedback \cite{wiseman_fb,Hofmann_qcontrol,Kor-01,Smith-Wiseman-2002,Gillett-2010, haroche_fb}. Here, we report the first implementation of quantum feedback control in a solid state system, in our case a superconducting quantum bit (qubit) coupled to a microwave cavity \cite{cQEDtheory}. Probing the state of the cavity with less than one
photon on average, implements a weak measurement of the qubit state. These photons are then directed to a high-bandwidth quantum-noise-limited amplifier \cite{hatridge_paramp,Vijay_qjumps}, which enables real-time monitoring of the state of the cavity\textemdash and hence that of the qubit\textemdash with high fidelity. We demonstrate quantum feedback control by inhibiting the decay of Rabi oscillations, allowing them to persist indefinitely \cite{korotkov_dir_fb}. This new ability permits active suppression of decoherence and defines a path for continuous quantum error correction \cite{qerrcorr,parity_cQED}. Other novel avenues include quantum state stabilization \cite{Hofmann_qcontrol,state_stable,Gillett-2010}, entanglement generation using measurement \cite{Korotkov_entang}, state purification \cite{state_pur_jacobs}, and adaptive measurements \cite{jacobs_adaptive_msmt,cook-2007}.

Feedback protocols in classical systems, from anti-lock brakes to pacemakers, use the outcome of a measurement to stabilize the system about a desired state. The operation of such feedback protocols is predicated on the idea that measurement does not \emph{itself} alter the state of the system. This is no longer true in quantum mechanics where measurement is necessarily invasive \cite{qmcontrol_book}.  In the Copenhagen interpretation, a quantum object can exist simultaneously in more than one stable state or eigenstate until observed\textemdash Schr\"{o}dinger's celebrated ``dead and alive" cat being the quintessential hypothetical example \cite{schrodcat}.  A sense of reality is restored by the act of measurement which forces the system ``instantaneously" into one of its eigenstates in a probabilistic fashion (the so-called measurement back-action). In light of this effect, how does one apply measurement-based feedback to a quantum system such as a qubit?  

One solution is to use weak measurements \cite{qmcontrol_book,qnoise_book} where we deliberately limit the rate ($\Gamma_{\mathrm{meas}}$) at which information is extracted, thereby slowing down the qubit's random walk towards an eigenstate. Integral to this scheme is a detector with efficiency $\eta_{\mathrm{det}} = \Gamma_{\mathrm{meas}}/\Gamma_{\varphi} \rightarrow 1$, where $\Gamma_{\varphi}$ is the ensemble averaged dephasing rate due to measurement back-action\cite{clerk_revmod}. The high detector efficiency allows us to faithfully track the qubit continuously, and steer it to a desired state using real-time feedback.

The experimental setup is shown in Fig. 1. Our quantum system (Fig. 1b) is an anharmonic oscillator realized by a capacitively-shunted Josephson junction, dispersively coupled to a 3D microwave cavity \cite{Paik_3DT}. We use its two lowest energy levels ($\omega _{01}/2\pi =5.4853$ GHz) to form a qubit (transmon \cite{transmontheory}). The cavity resonant frequency with the qubit in the ground state is $\omega _{c}/2\pi =7.2756$ GHz. The strongly coupled output port sets the cavity linewidth $\kappa/2\pi=13.4$ MHz while control and measurement signals are injected via the weakly coupled input port (Figs. 1a and 1b). The qubit-cavity coupling results in a state-dependent phase shift [$\Delta\phi =2\tan^{-1}{(2\chi/\kappa)}= 12^\circ, \, \chi/2\pi=0.687$ MHz] of the cavity output field \cite{cQEDtheory, wallraff_visibility}, with the state information contained in one quadrature of the signal. The cavity output is sent to a near-noiseless ($\eta_{\mathrm{det}} \sim 1$) phase-sensitive parametric amplifier (paramp) \cite{hatridge_paramp,Vijay_qjumps} which boosts the relevant quadrature to a level compatible with classical circuitry. The paramp output is further amplified and homodyne detected (Fig. 1c) such that the amplified quadrature (Q) contains the final measurement signal.

We acquire Rabi oscillations with the cavity continuously excited at $\omega_{\mathrm{r}}/2\pi=7.2749$ GHz $(\omega_{\mathrm{r}} \approx \omega_{\mathrm{c}} - \chi)$ with a mean cavity photon occupation $\bar{n}$ which controls the measurement strength (see section II of supplementary information for calibration of $\bar{n}$ ). The Rabi drive at the ac Stark shifted \cite{schusteracstark} qubit frequency ($\omega_{\mathrm 01} - 2 \chi \bar{n}$) is turned on for a fixed duration $\tau_{m}$. The amplitude is adjusted to yield a Rabi frequency $\Omega_{R}/2\pi=3$ MHz. First, we average $10^4$ measurement traces to obtain a conventional ensemble-averaged Rabi oscillation trace (Fig. 2a). Even though the qubit is continuously oscillating between its ground and excited states, the oscillation phase diffuses primarily due to measurement back-action. As a result, the averaged oscillation amplitude decays over time. The persistent nature of these oscillations, however, is evident in the frequency domain response \cite{saclay_bell_ineq}. We Fourier transform the individual measurement traces and plot the averaged spectrum in Fig. 2b (blue trace). A peak centered at 3 MHz with a width $\Gamma$ is observed and remains unchanged even when $\tau_{m}$ is much longer than the decay time of the ensemble averaged oscillations. A plot of the peak width $\Gamma$ for different measurement strengths (in units of $\bar{n}$) is shown in Fig. 2c.  As expected in the dispersive regime, $\Gamma$ and $\bar{n}$ are linearly related \cite{schusteracstark}. The vertical offset is dominated by pure enviromental dephasing $\Gamma_{\mathrm{env}}$ but has contributions from qubit relaxation ($T_{1}$) and thermal excitation into higher qubit levels; more details can be found in sections II and IV(C) of the supplementary information.

The ratio of the height of the Rabi spectral peak to the height of the noise floor has a theoretical maximum value of four \cite{Korotkov_p2p}, corresponding to an ideal measurement with overall efficiency $\eta =1$. For our setup, this efficiency can be separated into two contributions as $\eta = \eta_{\mathrm{det}} \ \eta_{\mathrm{env}}$. The detector efficiency is given by $\eta_{\mathrm{det}}=[1+ 2\, n_{\mathrm{add}}]^{-1}$ with $n_{\mathrm{add}}$ being the number of noise photons added by the amplification chain. The added noise is referenced to the output of the cavity and includes the effect of signal attenuation between the cavity and the amplifier. The effect of environmental dephasing $\Gamma_{\mathrm{env}}$ is modeled using $\eta_{\mathrm{env}} = (1+\Gamma_{\mathrm{env}} / \Gamma_{\varphi})^{-1}$. The best measurement efficiency we obtain experimentally is $\eta = 0.40$, with $\eta_{\rm det}=0.46$, and $\eta_{\rm env}=0.87$; more details can be found in section III of supplementary information.

We now discuss the feedback protocol which can be understood as a phase-locked loop stabilizing a quantum oscillator. The amplified quadrature (Q) is multiplied by a Rabi reference signal with frequency $\Omega_0 /2\pi =3$ MHz using an analog multiplier (Fig. 1d). The output of this multiplier is low-pass filtered and yields a signal proportional to the sine of the phase difference $\theta_{\mathrm{err}}$ between the 3 MHz reference and the 3 MHz component of the amplified quadrature. This ``phase error'' signal is fed back to control the Rabi frequency $\Omega_{\mathrm{R}}$ by modulating the Rabi drive strength with an upconverting IQ mixer (Fig. 1a). The amplitude of the reference signal controls the dimensionless gain $F$ given by $\Omega_{\mathrm{fb}} / \Omega_{\mathrm{R}} = -F\sin(\theta_{\mathrm{err}})$ where $\Omega_{\mathrm{fb}}$ is the change in Rabi frequency due to feedback. Fig. 2d shows a feedback-stabilized oscillation which persists for much longer than the original oscillation in Fig. 2a. In fact, within the limits imposed by our maximum data acquisition time of 20 ms, these oscillations persist indefinitely. The red trace in Fig. 2b shows the corresponding averaged spectra. The needle-like peak at 3 MHz is the signature of the stabilized Rabi oscillations.

To confirm the quantum nature of the feedback-stabilized oscillations, we perform state tomography on the qubit \cite{Steffen_tomo}.  We stabilize the dynamical qubit state, stop the feedback and Rabi driving after a fixed time (80 $\mu$s + $\tau_{\mathrm{tomo}}$ after starting the Rabi drive), and then measure the projection of the quantum state along one of three orthogonal axes.  This is done by using strong measurements (by increasing $\bar{n}$) with single-shot fidelity \cite{Vijay_qjumps, johnson-herald,dicarlo-init} which allow us to remove any data points where the qubit is found in the second excited state (see section IV(C) of supplementary information). By repeating this many times, we can determine $\langle \sigma_{X}\rangle$, $\langle\sigma_{Y}\rangle$, and $\langle\sigma_{Z}\rangle$, the three components of the Bloch vector for the ensemble qubit state.  Fig. 3a shows a plot of the Bloch vector components for different time points ($\tau_{\mathrm{tomo}}$) over one oscillation period [$1/(2\pi \Omega_0)$]. The Y and Z components are well fit by a sinusoidal function, whereas the X component is nearly zero as expected for a coherent Rabi oscillation about the X axis. The efficiency of the feedback process is reflected in the non-unit amplitude of these oscillations. This feedback efficiency $D$ is given by the time-averaged scalar product of the desired and actual state vectors on the Bloch sphere (see section IV(A) of supplementary information). In our experiment, the measurement is weak enough that the stabilized oscillations are sinusoidal and $D$ is approximately equal to the amplitude of these oscillations. 

In Fig. 3c we plot $D$ (red squares) versus the dimensionless feedback strength $F$. We find a maximum value of $D=0.45$ for the optimal choice of $F$. The dashed black line is a plot of the theoretical expression for $D$ given by
\begin{equation}
\label{deqn}
D=\frac{2}{\frac{1}{\eta }\frac{F}{\Gamma /\Omega _{R}}+\frac{\Gamma /\Omega _{R}}{F}}\,,
\end{equation}
which is derived using a simple analytical theory based on the Bayesian formalism for the qubit state trajectory (see section IV(A) of supplementary information) but does not account for finite feedback bandwidth, loop delays in the circuit or qubit relaxation.  The maximum value $D_{\mathrm{max}}=\sqrt{\eta}$ is obtained for an optimal feedback strength $F_{\mathrm{opt}}=\sqrt{\eta} \ \Gamma/\Omega_{R}$.  A value of $D_{\mathrm{max}}<1$ implies that the stabilized state is a mixed state; this occurs for $\eta<1$, implying that we have incomplete information about the qubit state. Another way to visualize the stabilized state with $D<1$ is shown in Fig. 3b. The Bloch vector during a single measurement is roughly within a certain angle of the desired state resulting in the averaged Bloch vector (red arrow) having a magnitude smaller than one. In principle, it is possible to approach a pure state with $D=1$ by ensuring $\eta=1$ and minimizing feedback loop delay. To account for the finite loop delay (250 ns), feedback bandwidth (10 MHz) and qubit relaxation ($T_{1}=20 \ \mu$s), we performed full numerical simulations of the Bayesian equations for qubit evolution (see section IV(B) of supplementary information). The results are shown as a black solid line in Fig. 3b and agree well with our experimental data. 

In conclusion, we have demonstrated a continuous analog feedback scheme to stabilize Rabi oscillations in a superconducting qubit, enabling them to persist indefinitely. The efficiency of the feedback is limited primarily by signal attenuation and loop delay, and can be improved in the near future with the development of on-chip paramps and cryogenic electronics, respectively. We anticipate that our present technology, with minimal modifications, is sufficient to implement continuous quantum error correction of a logical qubit encoded in multiple physical qubits using pairwise parity measurements \cite{qerrcorr,parity_cQED}. This development will also usher in a new era of measurement based quantum control for solid-state quantum information processing \cite{Hofmann_qcontrol,state_stable,Gillett-2010,Korotkov_entang,state_pur_jacobs,jacobs_adaptive_msmt,cook-2007}. 

We thank M. Sarovar for useful discussions and Z. Minev for assistance with numerical simulations. This research was supported in part (R.V., C.M., and I.S.) by the U.S. Army Research Office (W911NF-11-1-0029) and the Office of the Director of National Intelligence (ODNI), Intelligence Advanced Research Projects Activity (IARPA), through the Army Research Office (K.W.M., S.J.W and A.N.K.). All statements of fact, opinion or conclusions contained herein are those of the authors and should not be construed as representing the official views or policies of IARPA, the ODNI, or the US Government. D.H.S. acknowledges support from a Hertz Foundation Fellowship endowed by Big George Ventures.

\bigskip

\begin{figure*}[t]
\begin{center}
\includegraphics[width=80mm]{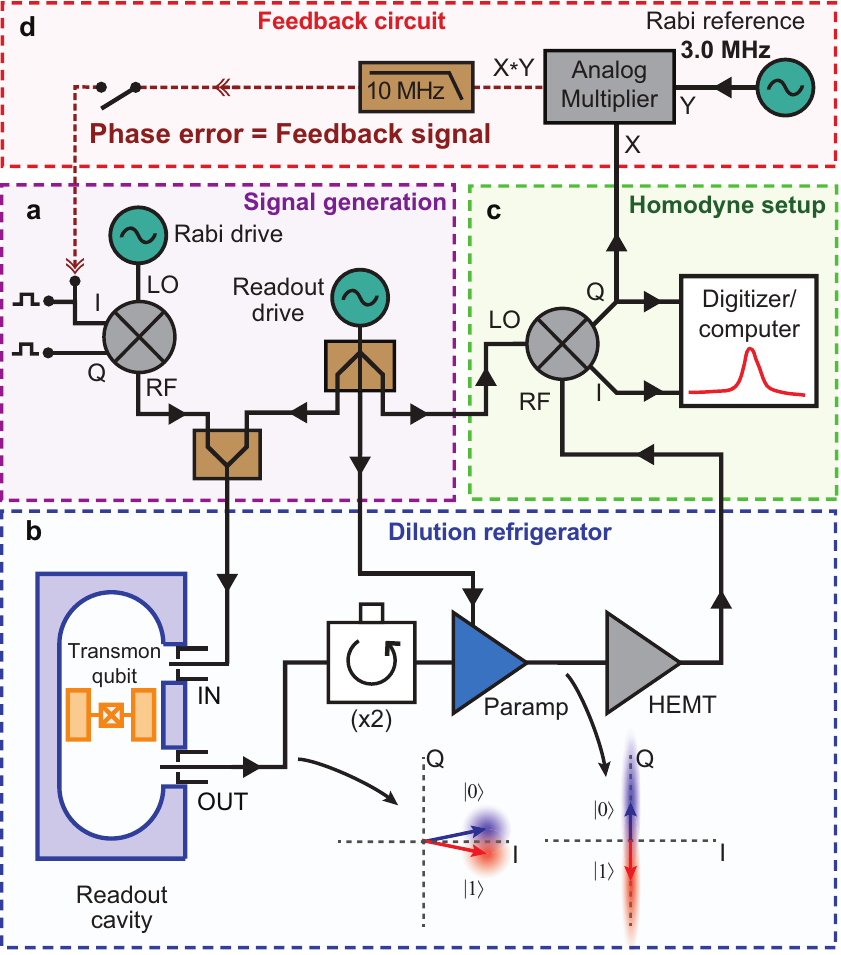}
\end{center}
\caption{
{\bf Experimental setup}. {\bf (a)} shows the signal generation setup.  One generator provides the Rabi drive at the ac Stark shifted qubit frequency ($\omega_{\mathrm 01} - 2 \chi \bar{n}$), while the output of another generator at 7.2749 GHz is split to create the measurement signal, paramp drive and local oscillator. The relative amplitudes and phases of these three signals are controlled by variable attenuators and phase shifters (not shown).  {\bf (b)} shows a simplified version of the cryogenic part of the experiment; all components are at 30 mK (except for the HEMT amplifier, which is at 4 K). The combined qubit and measurement signals enter the weakly coupled cavity port, interact with the qubit, and leave from the strongly coupled port. The output passes through two isolators (which protect the qubit from the strong paramp drive), is amplified, and then continues to the demodulation setup. The coherent state at the output of the cavity for the ground and excited states is shown schematically before and after parametric amplification. {\bf (c)} The amplified signal is homodyne detected and the two quadratures are digitized.  The amplified quadrature (Q) is split off and sent to the feedback circuit {\bf (d)}, where it is multiplied with the Rabi reference signal.  The product is low-pass filtered and fed back to the IQ mixer in {\bf (a)} to modulate the Rabi drive amplitude.}
\end{figure*}

\begin{figure*}[b]
\begin{center}
\includegraphics{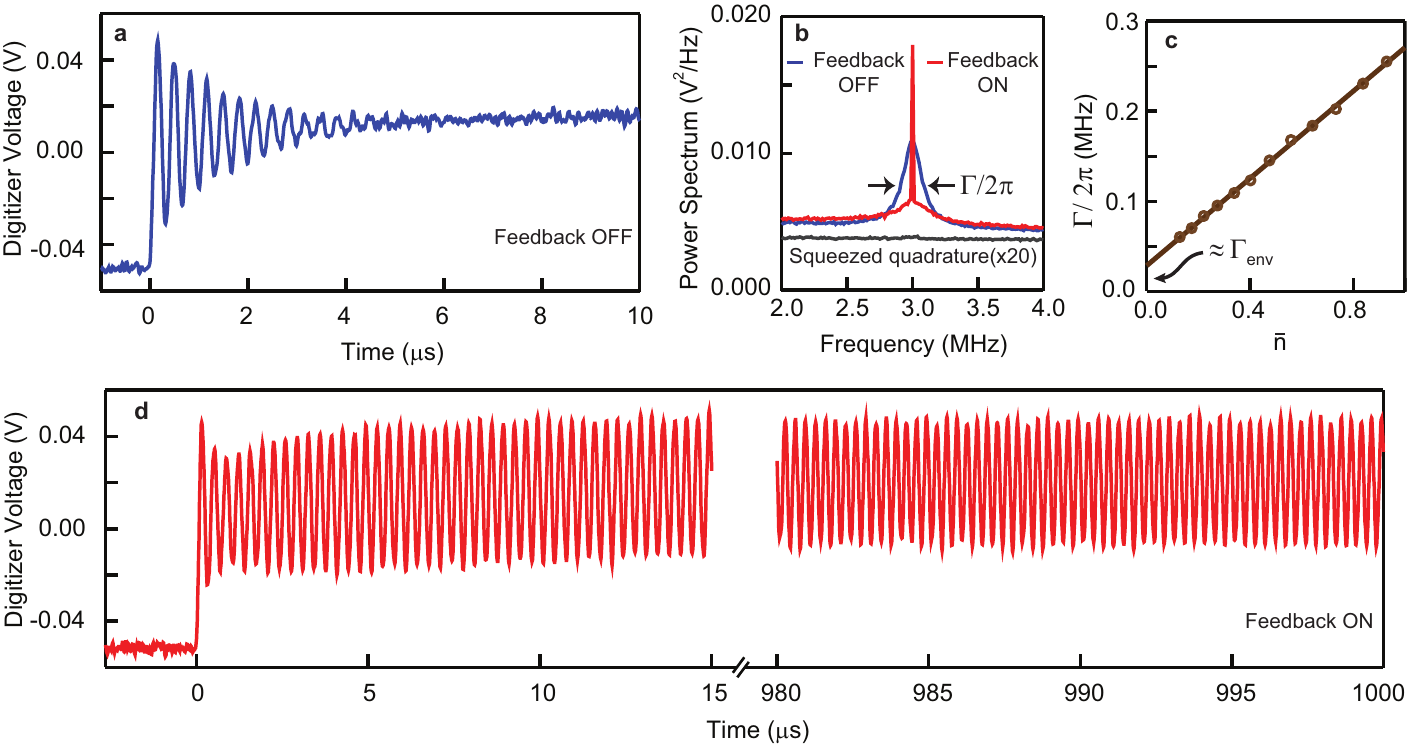}
\end{center}
\caption{
{\bf Rabi oscillations and feedback}. {\bf (a)} shows conventional ensemble-averaged Rabi oscillations measured using weak continuous measurement, which decay in time due to ensemble dephasing. In {\bf (b)}, the individual measurement traces from {\bf (a)} are Fourier transformed before averaging.  The averaged spectrum shows a peak at the Rabi frequency (blue trace) with a width $\Gamma /2\pi$ (FWHH). The grey trace shows an identically prepared spectrum for the squeezed quadrature (multiplied by 20 for clarity), which contains no qubit state information. $\Gamma$ is plotted as a function of cavity photon occupation $\bar{n}$ (measurement strength) in {\bf (c)}, showing the expected linear dependence. {\bf (d)} shows feedback-stabilized ensemble averaged Rabi oscillations, which persist for much longer times than those without feedback seen in {\bf (a)}. The corresponding spectrum, shown in {\bf (b)}, has a needle-like peak at the Rabi reference frequency (red trace). The slowly changing mean level in the Rabi oscillation traces in {\bf (a)} and {\bf (d)} is due to the thermal transfer of population into the second excited state of the qubit. See section IV(C) of supplementary information for more details.}
\end{figure*}

\begin{figure*}[htbp]
\begin{center}
\includegraphics{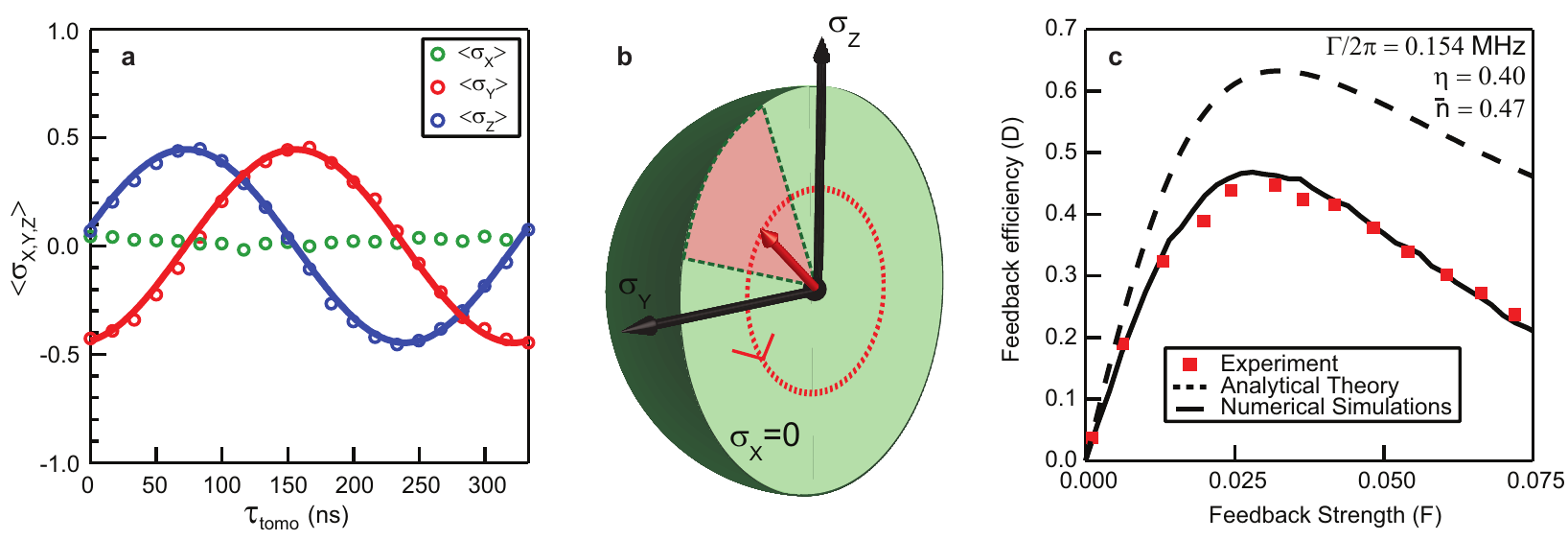}
\end{center}
\caption{
{\bf Tomography and feedback efficiency}. {\bf (a)} shows quantum state tomography of the feedback-stabilized state. We plot $\langle\sigma_{X}\rangle$ , $\langle\sigma_{Y}\rangle$ and $\langle\sigma_{Z}\rangle$ for different time points $\tau_{\mathrm{tomo}}$ in one full Rabi oscillation of the qubit. The solid lines are sinusoidal fits. The magnitude of these sinusoidal oscillations is approximately equal to the feedback efficiency $D=0.45$. {\bf (b)} shows a visualization of the feedback stabilized Bloch vector for $D<1$.  The Bloch vector during a single measurement remains roughly within a certain angle (shaded pie region) of the desired state; higher the efficiency $D$, smaller the angle. Tomography of the stabilized state measures the average over many iterations which reduces the length of the averaged Bloch vector (red arrow). In {\bf (c)}, we plot $D$ as a function of the dimensionless feedback strength $F$. Solid red squares are experimental data with a maximum value of $D=0.45$.  The dashed black line is a plot of Eq. (\ref{deqn}) with $\eta=0.40$ and $\Gamma/2\pi=0.154$ MHz ($\bar{n}=0.47$, $\Gamma_{\mathrm{ env}}/2\pi=0.020$ MHz), while the solid black line is obtained from full numerical simulations of the Bayesian equations including finite loop delay (250 ns) and feedback bandwidth (10 MHz).}
\end{figure*}

\end{document}


\title{ \ \ \ \ \ \ \ \ \ \underline{ SUPPLEMENTARY INFORMATION} \newline \newline Quantum feedback control of a superconducting qubit: Persistent Rabi oscillations}
\author{R. Vijay$^1$, C. Macklin$^1$, D. H. Slichter$^1$, S. J. Weber$^1$, K. W. Murch$^1$, R. Naik$^1$, \\
 A. N. Korotkov$^2$,  I. Siddiqi$^{1}$}
\affiliation{$^1$Quantum Nanoelectronics Laboratory, Department of Physics, University of California, Berkeley CA 94720}
\affiliation{$^2$Department of Electrical Engineering, University of California, Riverside, CA 92521}

\date{\today}

\maketitle

\date{\today}
\renewcommand{\thefigure}{S\arabic{figure}}

\section{Device fabrication and parameters}  
The transmon qubit was fabricated on a bare high-resistivity Si wafer using electron beam lithography and double-angle Aluminum evaporation with an intervening oxidation step. The qubit is a single Josephson junction with two rectangular paddles ($425\,\mu\rm{m} \times 225\, \mu\rm{m}$) which provide the shunting capacitance and coupling to the cavity. The cavity was machined out of Aluminum (6061 alloy). The quality factor of the cavity was adjusted by controlling the length of the center conductor of the SMA coaxial connector protruding into the cavity volume. This was done for the output port to provide strong coupling while the input port was coupled weakly to provide a net power transmission at resonance of -20 to -30 dB.
  
The Josephson ($E_J$) and charging energy ($E_C$) of the transmon qubit were determined by qubit spectroscopy which yielded transition frequencies $\omega _{01}/2\pi =5.4853$ GHz and $\omega _{02}/2\pi =10.7382$ GHz. We then calculated $E_J=19.274$ GHz and $E_C=0.211$ GHz giving $E_J/E_C = 91$. The qubit relaxation time was measured to be $T_1=20 \ \mu s$ while echo experiments yielded $T_2^\ast=8 \ \mu s$.

\section{Dispersive shift and photon number calibration}

In order to determine the dispersive shift $2\chi$ of the cavity between the qubit's ground and excited states, we use a combination of ac stark shift ($\Delta \omega_{ac} = 2\chi \bar{n}$) and measurement induced dephasing of the qubit ($\Gamma_{\varphi} = 8\chi^2 \bar{n}/\kappa  $), where $\bar{n}$ is the average photon occupation of the readout cavity \cite{schusteracstark}. To measure these quantities precisely, we perform a Ramsey fringe experiment where the free evolution period between the two $\pi / 2$ pulses is modified by exciting the cavity with a fixed power $\bar{P}$ at the readout frequency $\omega_{\rm r}$. By fitting the Ramsey fringes to an exponentially decaying sinusoidal function, we measure $\Delta \omega_{ac}$ by extracting the Ramsey frequency and the dephasing rate ($\Gamma_{\mathrm{Ramsey}} = \Gamma_{\varphi}  + \Gamma^\ast_{2}$) by extracting the decay constant. Here $T^\ast_{2} = 1/\Gamma^\ast_{2}$ is the decay constant of the Ramsey fringes in the absence of any photons in the cavity. This technique is significantly faster than conventional spectroscopy \cite{schusteracstark} and provides better precision in extracting $\Delta \omega_{ac}$  and $\Gamma_{\varphi}$. We repeat this process for different $\bar{P}$ and since $\bar{n} \propto \bar{P}$, a plot of $\Delta \omega_{ac}$ vs $\bar{P}$ and $\Gamma_{\mathrm{Ramsey}}$ vs $\bar{P}$ gives two straight lines with slopes $m_{ac}$ and $m_{\varphi}$ (Fig. S1). The ratio $m_{\varphi}/m_{ac} = 4\chi/\kappa$ then allows us to determine the dispersive shift $2\chi = 1.375$ MHz. We use this value of $2\chi$ and the stark shift data to get a calibration for the average photon number $\bar{n}$ in the cavity.

\section{Detector and measurement efficiency}
\label{sec_efficiency}

In our experiment, the overall measurement efficiency relevant for feedback is given by $\eta = \eta_{\mathrm{det}} \ \eta_{\mathrm{env}}$. The first term $\eta_{\mathrm{det}}$ accounts for the noise added by the amplification chain. To measure the noise added by the amplifier, one typically uses a calibrated noise source. Instead, we use the qubit+cavity system as a calibrated signal source. As discussed in the previous section, we can excite the cavity with a precise average photon occupation $\bar{n}$. This corresponds to a power radiated from the cavity $P_{\mathrm{rad}}= \hbar\omega_{r} \kappa$ where $\omega_{r}$ is the frequency of excitation. We send this signal to the paramp and measure the signal to noise ratio (SNR) at the output. This allows us to extract the noise floor of the amplification chain which includes a dominant contribution from the paramp and a smaller contribution from the HEMT amplifier. 

The noise floor referred to the input of amplification chain is given by $P_{\rm n} = \hbar \omega_{r} B/\eta_{\rm det}$ where B is the integration bandwidth (in Hz).  We calculate the detector efficiency as
	\be
		\eta_{\rm det}=\frac{{\rm SNR}}{   \bar{n}} \frac{B }{ \kappa}.
	\ee
We repeat this experiment for a range of frequencies within the paramp bandwidth and extract an average detector efficiency $\eta_{\rm det}=0.46$.  Note that the actual power reaching the paramp is smaller due to signal attenuation between the cavity and the paramp. This attenuation reduces the overall efficiency and our measurement technique correctly measures the efficiency of the amplification chain starting from the output of the cavity. With the help of cryogenic switches we independently measure the signal attenuation to be roughly 2.5 dB implying that this mechanism is the dominant mechanism for reduction in detector efficiency. Future experiments will endeavor to minimize this signal attenuation.

 The final contribution to the overall measurement efficiency is due to environmental decoherence via pure dephasing. The efficiency $\eta_{\mathrm{env}} = (1+\Gamma_{\mathrm{env}} / \Gamma_{\varphi})^{-1}$ characterizes how much of the total dephasing is due to measurement. In principle one can keep increasing $\Gamma_{\varphi}$ by increasing $\bar{n}$ to improve this efficiency but there are two practical constraints. The measurement should be weak enough such that it is not projective on the timescale of the Rabi period $\Omega_{\rm R}^{-1}$ thus leaving the qubit evolution oscillatory. Furthermore, the feedback bandwidth required increases with $\Gamma_{\varphi}$. Since the effective feedback bandwidth is fixed by the measurement chain, the feedback efficiency $D$ decreases with increasing measurement strength. The data shown in Fig. 2 and Fig. 3 corresponds to an optimal choice of $\Gamma_{\varphi} = 8\chi^2 \bar{n}/\kappa =0.134$ MHz to give the maximum value of $D$. Now dephasing due to low frequency noise does not affect the feedback efficiency because the system can track any slow variations in the qubit frequency (and consequently in the Rabi frequency). Hence, we set $\Gamma_{\mathrm{env}} = 1/T_2^\ast$ measured from echo experiments giving us $\Gamma_{\mathrm{env}}/2\pi = 0.02$ MHz, and $\eta_{\mathrm{env}}=0.87$. Note that this definition includes the dephasing contribution from  qubit relaxation ($\Gamma_1 / 2$).

\section{Bayesian formalism and feedback efficiency}
\label{sec_Bayes}

In this section, we will briefly describe the ``quantum Bayesian'' formalism
\cite{Kor-01}, which, broadly speaking, is similar to the ``quantum
trajectory'' theory
\cite{Wiseman-Milburn-QT,wiseman_fb,gambettapra2008}. The
assumptions needed for applicability of the formalism to a circuit
QED setup \cite{korotkov_cQED} (dispersive interaction, ``bad cavity''
regime, and weak response) are well satisfied in our experiment.
Since we use a phase-sensitive detector to amplify the
optimal signal quadrature, there is essentially no back-action on the qubit
from photon number fluctuations in the resonator \cite{korotkov_cQED},
and thus the present case is identical to that of a qubit measured 
by a quantum point contact. Note that we will be using the symbol $I(t)$ to describe the measurement output signal to be consistent with
reference \cite{Kor-01}. This should not be confused with the `I' quadrature of the mixed down signal discussed in the main text where the measurement output is the `Q' quadrature.

We will first consider the case where the detector is ideal ($\eta_{\rm det}=1$). The qubit evolution during the process of continuous measurement can be described using stochastic equations \cite{Kor-01}
for the qubit density matrix $\rho$. For a resonant Rabi drive, the equations (in Stratonovich form) are given by
\begin{equation}
 \dot{\rho}_{11}=-\dot{\rho}_{00}=-\Omega _{\mathrm{R}}\mathrm{Im}\rho
_{01}+\rho _{11}\rho _{00}\frac{2\Delta I}{S_{\mathrm{id}}}[I(t)-\frac{I_0+I_1}{2}]-\Gamma_{1} \rho_{11}  \label{Bayes_11}
\end{equation}
\begin{equation}
\dot{\rho}_{01} =i\frac{\Omega _{\mathrm{R}}}{2}\left( \rho _{11}-\rho
_{00}\right) -\frac{\Delta I}{S_{\mathrm{id}}}\rho _{01}\left( \rho _{11}-\rho
_{00}\right) [I(t)-\frac{I_0+I_1}{2}] -(\Gamma _{\mathrm{env}} + \frac{\Gamma_{1}}{2})\rho _{12}  \label{Bayes_01}
\end{equation}
where $\Omega_{\mathrm{R}}$ is the Rabi frequency, $\Gamma _{\mathrm{env}}$ is the environmental dephasing rate, $\Gamma_{1} = 1/T_{1}$ is the qubit relaxation rate. The measurement output signal $I(t)$ is given by
	\be
	I(t) = \frac{I_{0}+I_{1}}{2} + \frac{\Delta I}{2} [\rho_{11}-\rho_{00}] + \xi_{\mathrm{id}}(t)
	\label{meas_I}
	\ee
where $I_{0}$ and $I_{1}$ are the average output signals for the qubit in ground and excited state respectively, $\Delta I = I_{1}-I_{0}$ is the response and $\xi_{\mathrm{id}}$ is the white noise of an ideal detector characterized by the (one-sided) spectral density $S_{\mathrm{id}}$. Note that for solving these equations, the choice of $I_{0}$, $I_{1}$ and $S_{\mathrm{id}}$ is somewhat arbitrary and it is the ratio of $(\Delta I)^{2}$ and $S_{\mathrm{id}}$ that matters. The strength of the measurement which is characterized by the measurement induced dephasing rate is given by
	\be
	\Gamma_{\varphi}=\frac{(\Delta I)^{2}}{4S_{\mathrm{id}}} = \frac{8\chi^{2} \bar{n}}{\kappa}
	\ee

We use two slightly different methods to account for measurement efficiency $\eta$ when solving equations (\ref{Bayes_11}), (\ref{Bayes_01}) and (\ref{meas_I}) analytically (section \ref{D_analytic}) or numerically (section \ref{D_numeric}).

\subsection{Analytical derivation of feedback efficiency}
\label{D_analytic}

We now derive an analytical expression for the
feedback efficiency $D$, neglecting delays in the feedback loop and
the effect of finite bandwidth in the resonator, amplifier,
and feedback circuit. We also assume weak
coupling, and no qubit energy relaxation ($\Gamma_{1}=0$). We do, however, take into account
arbitrary measurement efficiency $\eta$. The total dephasing rate is given by $\Gamma = \Gamma_{\varphi}+\Gamma_{\rm env}$.

To obtain a closed form expression, it is possible to
reduce the number of qubit degrees of freedom down to only one.
First, on account of a resonant Rabi drive which rotates the qubit
about the $x$-axis, the process of measurement eventually attracts the
qubit to the $x=0$ plane. Second, in the absence of energy relaxation, we
can consider the qubit state as pure, ascribing any measurement 
inefficiency ($\eta<1$) to the additional noise at the detector
output \cite{Kor-nonideal}.  Note that $\eta$ includes the effect of environmental dephasing $\Gamma_{\rm env}$ (via $\eta_{\rm env}$) as well as detector efficiency $\eta_{\rm det}$. So we set $\Gamma_{\rm env}=0$ in equation (\ref{Bayes_01}) and model both environmental dephasing and detector inefficiency by adding the noise term $\xi_{\mathrm{add}}(t)$ to equation (\ref{meas_I}) and we get
	\be
	I(t) = \frac{I_{0}+I_{1}}{2} + \frac{\Delta I}{2} [\rho_{11}-\rho_{00}] + \xi_{\mathrm{id}}(t) + \xi_{\mathrm{add}}(t)
	\label{meas_I_analytic}
	\ee
where the added noise  $\xi_{\mathrm{add}}(t)$ has  a spectral density $S_{\mathrm{add}} = S_{\rm out} - S_{\rm id}$, where $S_{\rm id}=(\Delta I)^{2}/ (4\,\Gamma)$ and $S_{\rm out} = S_{\rm id}/\eta$ is the total output noise. Therefore, the qubit state evolution can
be described by only one parameter, the polar (zenith) angle $\theta
(t)$ on the Bloch sphere:
    \be
    z(t)= \cos [\theta(t)], \,\,\, y(t)=\sin[\theta(t)], \,\,\, x(t)=0.
    \ee

The goal of the feedback is to produce $\theta(t)=\Omega_0 t$ with a
fixed frequency $\Omega_0$. We characterize the feedback
efficiency \cite{korotkov_dir_fb} by
    \be
    D=\overline{\cos \theta_{\rm err}(t)}, \,\,\, \theta_{\rm err} = \theta-\Omega_0 t,
    \ee
which is the time-averaged scalar product of the desired and actual state vectors on the Bloch sphere. An equivalent definition via the
density matrix is $D=2\, \overline{{\rm Tr}(\rho_{\rm desired}
\rho_{\rm actual})}-1$. The qubit ``phase shift error'' $\theta_{\rm err}$ evolves as \cite{korotkov_dir_fb}
    \be
    \dot\theta_{\rm err} = -\frac{\Delta I}{S_{\rm id}} \sin\theta
\left(\frac{\Delta I}{2}\cos \theta +\xi_{\rm id}\right) + \Omega_{\rm
fb} (t),
    \label{evol}\ee
where $\Omega_{\rm fb}(t)$ is the modulated part of the Rabi
frequency due to feedback
    \be
\Omega_{\rm R}(t) =\Omega_0+\Omega_{\rm fb} (t),
   \label{fb0} \ee
 which is given by the ``direct feedback'' control law:
    \be
    \frac{\Omega_{\rm fb}(t)}{\Omega_0} = F \frac{4}{\Delta I} \,
    \sin (\Omega_0 t)
    \left( I(t-0)- \frac{I_0+I_1}{2} \right) .
    \label{fb}\ee
Here $F$ is the dimensionless feedback strength and the choice of the
normalization factor $4/\Delta I$ corresponds to $\Omega_{\rm
fb}/\Omega_0=-F\sin\theta_{\rm err}$ on average.  We use the notation $I(t-0)$
which reminds us about the unavoidable delay in the feedback loop and is assumed to be very small for this calculation. Moreoever,  we have neglected effects of finite feedback bandwidth.

The qubit evolution (\ref{evol}) is written in the Stratonovich
form; converting it into the It\^o form (for averaging) we obtain
the extra term $[(\Delta I)^2/4S_{\rm id}]\sin\theta\cos\theta$, which
comes from the measurement part of Eq.\ (\ref{evol}) [the feedback
(\ref{fb}) is the same in both forms, and there is no cross-term
because of  non-zero loop delay -- see Ref.\
\cite{wiseman_fb} and also Comment and Reply]. However, this
extra term in the It\^o form is not important because we average
the evolution of the phase shift $\theta_{\rm err}$ over the Rabi period. The
averaging is simple when $\theta_{\rm err}$ evolves slowly, so that $\theta_{\rm err}$
is uncorrelated with $\theta$. Thus, we need to assume weak
coupling, $\Gamma \ll \Omega_0$ and weak feedback, $F\ll 1$. Averaging cancels the product $\sin\theta\cos\theta$ and replaces $\sin(\Omega_0 t)\cos\theta$ with $-(\sin\theta_{\rm err} )/2$; thus we obtain
    \begin{eqnarray}
&&  \dot\theta_{\rm err} =
 \left( \frac{4F\Omega_0}{\Delta I} \sin(\Omega_0 t)
 -\frac{\Delta I}{S_{\rm id}} \sin (\Omega_0 t+\theta_{\rm err}) \right) \xi_{\rm id}
 \,\,
 \nonumber \\
&&\hspace{0.7cm} + \frac{4F\Omega_0}{\Delta I} \,
    \sin (\Omega_0 t) \,    \xi_{\rm add}  -F\Omega_0 \sin\theta_{\rm err}.
    \label{evol-main}\end{eqnarray}

    In this equation, the last term attracts the phase shift
$\theta_{\rm err}$ to zero, while the noise terms cause diffusion of
$\theta_{\rm err}$. Examining the term in large parentheses, it is clear why there is an optimum value of the feedback strength $F$. For example, for
an ideal detector ($\eta=1$, then $\xi_{\rm add}=0$), the effect of
the noise $\xi_{\rm id}$ can be compensated when $4F\Omega_0/\Delta
I = \Delta I/S_{\rm id}$, leading asymptotically to full
synchronization, $\theta_{\rm err}(t)=0$. This compensation was studied
previously \cite{Hofmann_qcontrol,wang_fb_TLA} in the context of qubit state
stabilization about a fixed point on the Bloch sphere.

    Now let us average the noise in Eq.\ (\ref{evol-main}) over a Rabi
period. We can replace $\sin (\Omega_0 t)\,\xi_{\rm add}$ with
$\tilde\xi_{\rm add}/\sqrt{2}$, where $\tilde\xi_{\rm add}$ is also
white noise with the same spectral density as $\xi_{\rm add}$.
Averaging the term with $\xi_{\rm id}$ is similar, but slightly
more cumbersome. We first rewrite it as $[A\cos (\Omega_0 t)+B\sin
(\Omega_0 t)]\xi_{\rm id}$ with $A=-(\Delta I/S_{\rm id})\sin
\theta_{\rm err}$ and $B=4F\Omega_0/\Delta I-(\Delta I/S_{\rm
id})\cos\theta_{\rm err}$. Averaging over a Rabi period then gives
$\sqrt{(A^2+B^2)/2} \, \tilde\xi_{\rm id}$ with a similar white
noise, $S_{\tilde{\xi}_{\rm id}}=S_{\rm id}$. Let us now add the
uncorrelated contributions from the noises $\tilde\xi_{\rm id}$ and
$\tilde\xi_{\rm add}$, and convert the result into a noise $C\,
\tilde\xi_{\rm out}$, where $\tilde\xi_{\rm out}$ has the same
spectral density $S_{\rm out}$ as the output noise and
$C^2=\eta(A^2+B^2)/2+(1-\eta)(4F\Omega_0/\Delta I)^2/2$. In this way
we replace Eq.\ (\ref{evol-main}) with
   \begin{eqnarray}
&&  \dot\theta_{\rm err} =  -F\Omega_0 \sin\theta_{\rm err} + C \, \tilde \xi_{\rm
out}, \,\,\,\, S_{\tilde \xi_{\rm out}}= S_{\rm out} ,
 \label{evol-eff}\\
&& C^2= \frac{2F\Omega_0}{S_{\rm out}} \left( \frac{1}{\eta}\,
\frac{F}{\Gamma/\Omega_0} +\frac{\Gamma/\Omega_0}{F}-2\cos\theta_{\rm err}
\right) .
    \label{C-def}\label{evol-3}\end{eqnarray}

    This is a Langevin equation, and the corresponding Fokker-Planck
equation for the probability distribution $P(\theta_{\rm err}, t)$ is
    \be
\frac{\partial P}{\partial t}= \frac{\partial (F\Omega_0\sin\theta_{\rm err}
\, P)} {\partial \theta_{\rm err}} + \frac{1}{4}\, \frac{\partial^2 (C^2
S_{\rm out} P)}{\partial \theta_{\rm err}^2},
    \label{F-P}  \ee
where $P(\theta_{\rm err})$ is $2\pi$ periodic and is normalized as
$\int_{-\pi}^{\pi} P(\theta_{\rm err})\,d\theta_{\rm err} =1$. The stationary solution
$P_{\rm st}(\theta_{\rm err})$ then satisfies equation $d(C^2 S_{\rm out}
P_{\rm st})/d\theta_{\rm err} + 4F\Omega_0\sin\theta_{\rm err}\, P_{\rm st} ={\rm
const}=0$, where the constant is zero because of the symmetry
between $\theta_{\rm err}$ and $-\theta_{\rm err}$. Using the $C^2(\theta_{\rm err})$ dependence
from Eq.\ (\ref{C-def}), we get
    \be
 P_{\rm st}(\theta_{\rm err})= p_0 \left(
\frac{1}{\eta}\, \frac{F}{\Gamma/\Omega_0}
+\frac{\Gamma/\Omega_0}{F}-2\cos\theta_{\rm err} \right) ^{-2},
    \label{P_st}\ee
where the normalization constant $p_0$ can also be calculated
explicitly (not needed here).

    Finally, from the stationary probability distribution for
the phase shift $\theta_{\rm err}$, we calculate the feedback efficiency as
$D=\int_{-\pi}^\pi \cos\theta_{\rm err}\, P_{\rm st}(\theta_{\rm err}) \, d\theta_{\rm err}$
and thus obtain the analytical formula,
    \be
    D=\frac{2}{\displaystyle \frac{1}{\eta}\, \frac{F}{\Gamma/\Omega_0}
+\frac{\Gamma/\Omega_0}{F}} ,
    \label{D-res}\ee
which we have also confirmed by numerical simulations described later. From Eq.\ (\ref{D-res}) it is straightforward to calculate the optimal value
of the feedback factor $F$ and corresponding maximum value for $D$:
    \be
    F_{\rm opt}=\sqrt{\eta} \, \frac{\Gamma}{\Omega_0} , \,\,\,
    D_{\rm max}=\sqrt{\eta}.
    \label{optimum}\ee

Notice that $F_{\rm opt}\ll 1$ for weak coupling,
$\Gamma\ll \Omega_0$, so the assumption of weak feedback,
$F\ll 1$, is satisfied. Also, 
$|\Omega_{\rm fb}(t)| \ll \Omega_0$ if the filter bandwidth is well
below $\Omega_0^2/\Gamma$, which is satisfied in the experiment.

\subsection{Numerical simulations}
\label{D_numeric}

We now discuss numerical simulations of the Bayesian equations. To avoid the complications with the Stratonovich vs the It\^o form for the stochastic differential equations, we used a two step process to evolve equations (\ref{Bayes_11}) and (\ref{Bayes_01}). First, we set $\Delta I = 0$ i.e. we suppress the measurement and evolve the the resulting ordinary differential equations using a $4^{\mathrm{th}}$ order Runge-Kutta step. We then include the effect of the measurement by performing a Bayesian \cite{Kor-01,korotkov_cQED} update which we describe below. The measurement output $I_{\mathrm{m}} = \tau^{-1} \int_{t}^{t+\tau } I(t^{'})dt^{'}$ in a given time interval $\tau$, is drawn from a Gaussian probability distribution with standard deviation $\sigma = \sqrt{S_{\rm id}/(2\tau)}$ and centered around $I_{0}$ and  $I_{1}$ for the qubit in state $|0\rangle$ and $|1\rangle$ respectively. The conditional probability distributions are given by
	\be
	P(I_{\mathrm{m}}\mid|0\rangle) = \frac{1}{\sqrt{2\pi \sigma}} \mathrm{exp}\left[-\frac{(I_{\mathrm{m}}-I_{0})^2}{2 \sigma^{2}}\right], \, \,\, \,  P(I_{\mathrm{m}}\mid|1\rangle) = \frac{1}{\sqrt{2\pi \sigma}} \mathrm{exp}\left[-\frac{(I_{\mathrm{m}}-I_{1})^2}{2 \sigma^{2}}\right]
	\ee 

Given an initial qubit state $\rho(t)$, the measurement outcome is drawn from a combined probability distribution 
	\be
	P(I_{\mathrm{m}}) =  \rho_{00}(t) P(I_{\mathrm{m}}\mid |0\rangle) +  \rho_{11}(t) P(I_{\mathrm{m}}\mid |1\rangle).
	\label{bayes_prob2}
	\ee
We use a combination of a binomial and a Gaussian random number generator to create a measurement outcome $I_{\mathrm{m}}$ which is then used to update the qubit state using the following equations \cite{Kor-01}
\begin{equation}
\rho_{11}(t+\tau) =\frac{\rho_{11}(t) P(I_{\mathrm{m}}\mid |1\rangle) }{P(I_{\mathrm{m}})}   , \, \, \,\,\, \rho_{00}(t+\tau) = \frac{\rho_{00}(t) P(I_{\mathrm{m}}\mid |0\rangle) }{P(I_{\mathrm{m}})}    \label{bu_11}
\end{equation}
\begin{equation}
\rho_{01}(t+\tau) = \rho_{01}(t)  \frac{\sqrt{\rho_{11}(t+\tau) \rho_{00}(t+\tau)}}{\sqrt{\rho_{11}(t) \rho_{00}(t)}}   \label{bu_01}
\end{equation}

This process is repeated for each time step to obtain the qubit density matrix $\rho(t)$ and the measurement output $I(t)$ as a function of time. If $\eta_{\rm det}<1$,  we add $\xi_{\mathrm{add}}(t)$ to $I(t)$ which is generated using an appropriate Gaussian white noise generator. Since $\Gamma_{\rm env}$ is now included in equation (\ref{Bayes_01}), the spectral density for the added noise is given by $S_{\mathrm{add}} = S_{\rm out} - S_{\rm id}$, where $S_{\rm out} = S_{\rm id}/\eta_{\rm det}$ is the total output noise. In other words, the extra noise only corresponds to detector inefficiency.

The output signal $I(t)$ is low pass filtered with a 10 MHz cutoff to account for the bandwidth of the paramp. To create the  feedback signal, we remove any dc offsets and multiply this output with the reference signal $\sin(\Omega_\mathrm{0}t)$, where $\Omega_{\mathrm{0}}/2\pi=3$ MHz.  We then implement feedback by modifying  $\Omega_{R} \rightarrow \Omega_{R}(t)$ in equations (\ref{Bayes_11}) and (\ref{Bayes_01}) using equations (\ref{fb0}) and (\ref{fb}). Feedback loop delay ($\tau_{\rm delay}$) is included by modifying the R.H.S of equation (\ref{fb}) so that $t \rightarrow t-\tau_{\rm delay}$
while feedback circuit bandwidth is included by filtering $\Omega_{\rm fb}(t)$ with a 10 MHz low-pass filter before adding it to equation (\ref{fb0}) . With the feedback modified qubit state $\rho(t)$ we can compute feedback efficiency $D$ as described in the previous section.

\subsection{Thermal fluctuations and higher qubit levels}
\label{subsec_thermal}

The discussion so far has assumed that the effective temperature of the qubit $T_{\rm qubit} \ll \hbar \omega_{\rm 01}/k_B$ where $\omega_{\rm 01}$ is the transition frequency between the ground and first excited state. Even though the dilution fridge temperature $T_{\rm}=30$ mK, we find significant thermal population of the first excited state corresponding to an effective temperature of approximately 140 mK. We believe that this is due improper thermalization of the qubit sample inside the Aluminum cavity. Further, the transmon qubit has higher levels with similar transition frequencies between neighbouring levels ($\omega_{\rm 12} \lesssim \omega_{\rm 01}$) and we observe few percent population in the second (and higher) excited states. 

We measure these populations using strong measurements which allows us to discriminate between the first 4 levels of the transmon with high single-shot fidelity \cite{Vijay_qjumps,johnson-herald,dicarlo-init}. Fig. S2 shows the single-shot histograms and one can clearly resolve four peaks in the distribution. These correspond to populations $P_0=0.83, P_1=0.13, P_2 = 0.03, P_{\rm 3+}=0.01$. Further, in the presence of Rabi driving (with or without feedback) the population in the 2nd excited state is enhanced up to 7.5 \% and 2.5 \% in higher levels. As mentioned in the main text, staet tomography of the stabilized state is performed using single-shot strong measurements and hence we can remove data points where we find the qubit state in the second excited state or higher. This only affects the value of D by about 10\%. For including these effects in the numerical simulations, we use a simple model described below. 

We consider only three levels (\kg, \ke, \kf) where \kf \ models all levels outside the \kg-\ke \ sub-space. Further, we only include the diagonal matrix element for \kf \  i.e. $\rho_{22}$ and all other terms involving the third level are set to zero. This is justified since we are only interested in the population of the higher levels and the off-diagonal terms representing the coherence are not important. Now the equations for the density matrix without the measurement terms ($\Delta I = 0$) are given by
\begin{equation}
\dot{\rho}_{00}=\Omega _{\mathrm{R}}\mathrm{Im}\rho
_{01}+\Gamma_{1} (\rho_{11}-\rho_{11,\rm st})  \label{Bayes_00a}
\end{equation}
\begin{equation}
 \dot{\rho}_{11}=-\Omega _{\mathrm{R}}\mathrm{Im}\rho
_{01}-\Gamma_{1} (\rho_{11}-\rho_{11,\rm st}) + \Gamma_{2} (\rho_{22}-\rho_{22,\rm st})   \label{Bayes_11a}
\end{equation}
\begin{equation}
\dot{\rho}_{01} =i\frac{\Omega _{\mathrm{R}}}{2}\left( \rho _{11}-\rho
_{00}\right)-\frac{1}{T_2^\ast} \rho _{12}  \label{Bayes_01a}
\end{equation}
\begin{equation}
 \dot{\rho}_{22}=-\Gamma_{2} (\rho_{22}-\rho_{22,\rm st})   \label{Bayes_22a}
\end{equation}
where in equation (\ref{Bayes_01a}) we have replaced $\Gamma _{\mathrm{env}} + \Gamma_{1}/2$ in equation (\ref{Bayes_01}) with $1/T_2^\ast$ as explained in section \ref{sec_efficiency}. Also $\rho_{11,\rm st}$ and $\rho_{22,\rm st}$ are the steady state thermal populations in state \ke \ and \kf \  respectively and we use the experimentally determined values. We ignore any relaxation from \kf \ directly to  \kg \ while the relaxation rate between \kf \ and \ke \ is modelled by $\Gamma_{2}$. We set $\Gamma_{2}= 2 \Gamma_1$ which is a good approximation for a weakly anharmonic qubit like the transmon. 

We now have a measurement current $I_2$ for the second excited state and we use $I_2 = I_1 + \Delta I$ which is a good approximation for a transmon in a cavity with $\chi \ll \kappa$ as is evident from Fig. S2. The Bayesian update for measurement now includes three levels. Equation (\ref{bayes_prob2}) now becomes
	\be
	P(I_{\mathrm{m}}) =  \rho_{00}(t) P(I_{\mathrm{m}}\mid |0\rangle) +  \rho_{11}(t) P(I_{\mathrm{m}}\mid |1\rangle)+ \rho_{22}(t) P(I_{\mathrm{m}}\mid |2\rangle)
	\label{bayes_prob3}
	\ee
where $P(I_{\mathrm{m}}\mid |0\rangle)$ and $P(I_{\mathrm{m}}\mid |1\rangle)$ are given by equation (\ref{bu_11}) and 
\be
	P(I_{\mathrm{m}}\mid|2\rangle) = \frac{1}{\sqrt{2\pi \sigma}} \mathrm{exp}\left[-\frac{(I_{\mathrm{m}}-I_{2})^2}{2 \sigma^{2}}\right]
\label{prob_state2}
	\ee 
is the conditional probability distribution for measurement of state \kf. Measurement outcomes $I_{\rm m}$ now use a combination of a trinomial (to choose one of the levels) and a Gaussian random number generator. Equations \ref{bu_11} and \ref{bu_01}  are still valid for the Bayesian update but we have a new equation for $\rho_{22}$ given by
\begin{equation}
\rho_{22}(t+\tau) =\frac{\rho_{22}(t) P(I_{\mathrm{m}}\mid |2\rangle) }{P(I_{\mathrm{m}})} . \label{bu_22}
\end{equation}

As in the experiment, we run the numerical simulation for $80 \ \mu s$ and keep only those iterations where the system remains in the \kg-\ke \ sub-space sub-space. Fig. S3 shows a typical qubit trajectory generated from numerical simulations. A thermally excited quantum jump into the second excited state which lasts for about $10 \ \mu s$ is clearly visible.

\newpage

\begin{figure}[t]
\begin{center}
\includegraphics{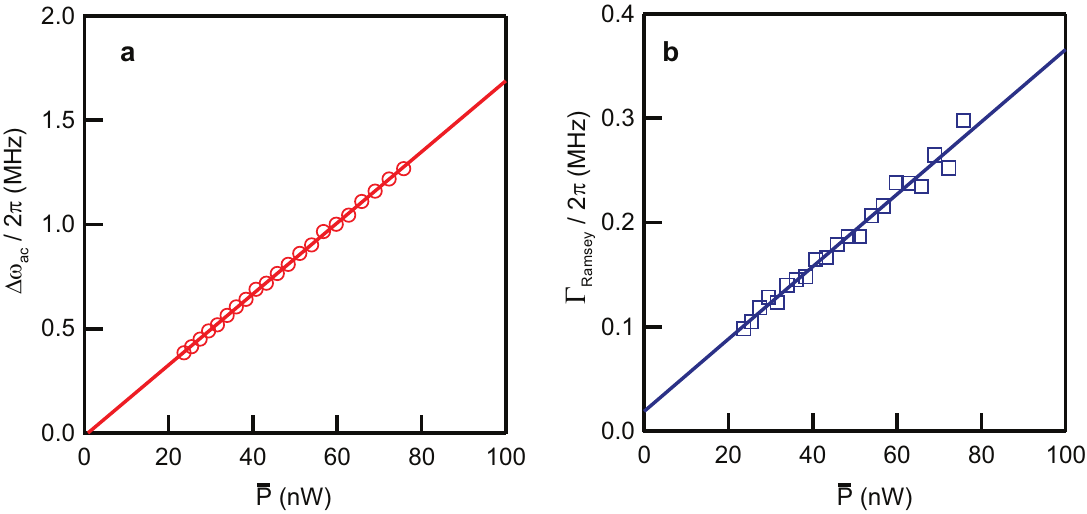}
\caption{
{\bf Dispersive shift and photon number calibration.} {\bf a} Ac stark shift $\Delta \omega_{ac}$ as a function of measurement drive power $\bar{P}$. {\bf b} Dephasing rate $\Gamma_{\mathrm{Ramsey}} = \Gamma_{\varphi}  + \Gamma_{2}$ as function $\bar{P}$.  Both quantities are extracted from Ramsey fringes by fitting an exponentially decaying sinusoidal function. The solid lines are linear fits with slopes $m_{ac}$ and $m_{\varphi}$ respectively. The power range corresponds to $\bar{n} < 1$.}
\end{center}
\end{figure}

\begin{figure}[htb]
\begin{center}
\includegraphics{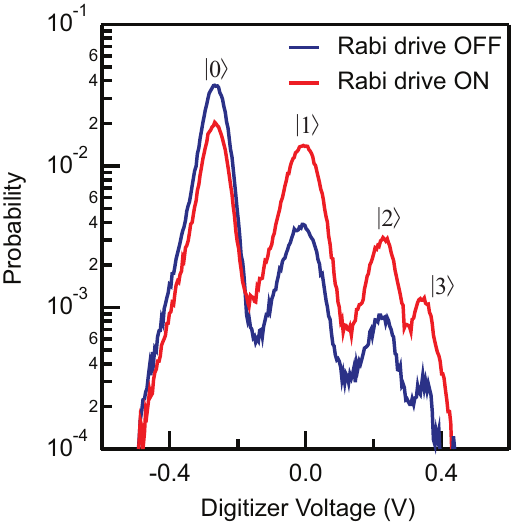}
\caption{
{\bf Thermal population.} The histograms are created from $3 \times 10^5$ single-shot measurements and cleary show four prominant peaks in the probability distribution corresponding to the first four levels of the transmon qubit. The blue trace corresponds to measurements made after letting the qubit relax to its steady state in the absence of any Rabi driving. The red trace corresponds to steady state with Rabi driving on the 0-1 transition and shows increased population in the second and higher excited states. }
\end{center}
\end{figure}

\begin{figure}[htb]
\begin{center}
\includegraphics{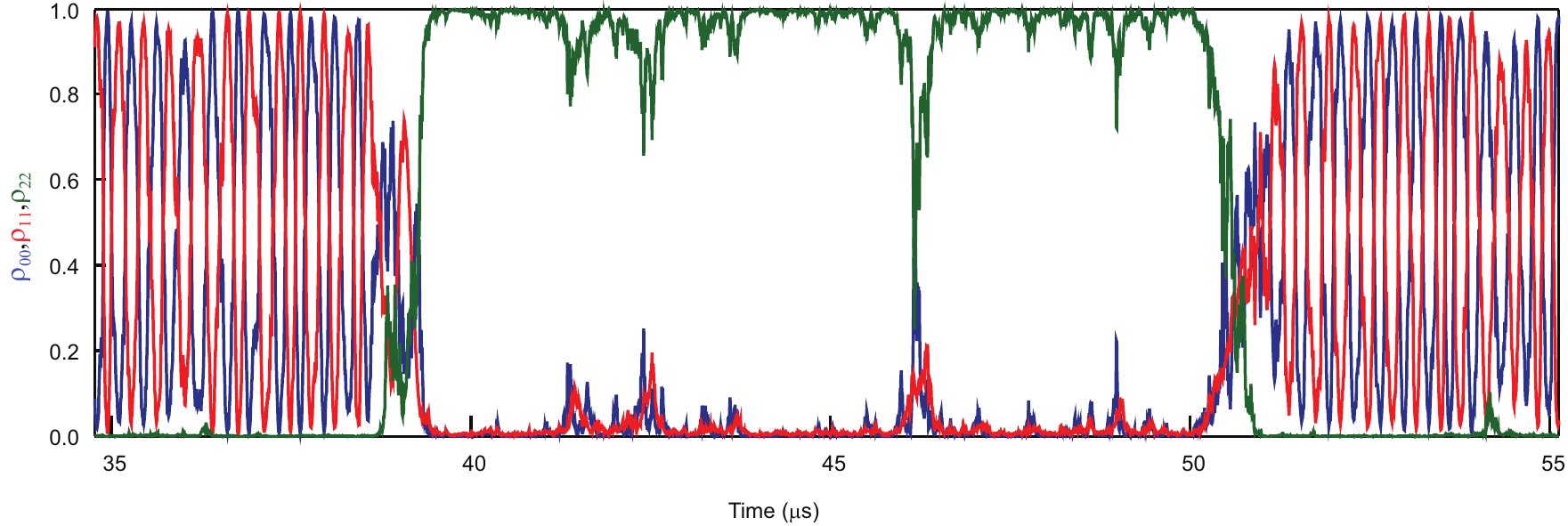}
\caption{
{\bf Simulated qubit trajectory}. The figure shows a typical qubit trajectory generated using numerical simulations and plots $\rho_{00}$ (blue) , $\rho_{11}$ (red) and $\rho_{22}$ (green). At the beginning, the qubit is undergoing stabilized Rabi oscillations which is interrupted due to a thermally excited quantum jump into the second excited state. The system remains in that state for about $10 \ \mu s$ before falling back into the 0-1 sub-space and the Rabi oscillations resume.}
\end{center}
\end{figure}